\newcommand{\Ms}{$M_{\odot}$}

\newcommand{\Zs}{$Z_{\odot}$}

\documentclass{prantzosjenam}
\usepackage{psfig}

\begin{document}

\runauthor{Prantzos}

\begin{frontmatter}

\title{Yields of Massive Stars and their Role\\
 in Galactic Chemical Evolution Studies}

\author[]{N. Prantzos}
\address{Institut d'Astrophysique de Paris, 98bis Bd. Arago, 75014 Paris}

\begin{abstract}
We review 
the yields of intermediate mass elements (from C to Zn) from
massive stars and their associated uncertainties, 
in the light of recent theoretical results.
We consider the role of those yields for our understanding
of the chemical evolution of the solar neighbourhood and of the
halo of our Galaxy. 
Current yields reproduce in a satisfactory way the solar system 
composition, but several problems remain concerning
 abundance ratios in halo stars.
\end{abstract}

\end{frontmatter}

\section{Introduction}

From the three main ingredients required to follow the chemical evolution of 
a galaxy,  only one, namely the stellar yields, can be calculated from first 
principles at present. The other two (star formation rate or SFR and stellar
initial mass function or IMF) can only be evaluated on empirical basis.

Massive stars are the main producers of most of the heavy isotopes in the
Universe.
Elements up to Ca are
mostly produced in such stars by hydrostatic burning, whereas Fe peak
elements are produced by the final supernova explosion (SNII), as well as
by white dwarfs exploding in binary systems as SNIa. Most of He, C, N and
minor CO isotopes, as well as s-nuclei comes from intermediate mass
stars (2-8 \Ms), which are not considered here. 

In Sec. 2 we discuss the various uncertainties still affecting the
yields of intermediate mass elements (between C and Zn) from massive stars.
In Sec. 3 we  analyse the successes and failures of current yields in 
reproducing the solar system elemental and isotopic composition.
In Sec. 4 we extend the investigation into the elemental composition of
stars of the Milky Way halo, formed more than $\sim$12 Gyr ago by
the ejecta of low metallicity stars; despite several successes, some recent 
observations cannot be interpreted in terms of currently available yields
and require a revision of our ideas on stellar nucleosynthesis.

\section {Yields of Massive Stars: Overview and Uncertainties}

Stars with a main sequence mass M$_{UP} > 8$ \Ms \ 
(or even lower, if convection
criteria leading to large convective cores are adopted) produce
at the end of their  hydrostatic evolution an Fe core, either by
quiescent  Si-burning (for $M > 11$ \Ms) or by electron captures
in a degenerate ONeMg core (for $M \sim 8$-11 \Ms). 
The structure and composition of the  ``onion-skin''
star at that stage reflects the
combined effect of (i) the various mixing mechanisms (convection,
semi-convection, rotational mixing etc.), determining the extent of the
various layers, (ii) the amount of mass-loss 
(affecting mostly the yields of the He and CNO nuclei, present in the
outer layers)  and (iii) the rates of the relevant nuclear ractions
(determining the abundances of the various species in each layer). 
Due to their
nuclear stability, $\alpha$-isotopes 
($^4$He, $^{12}$C, $^{16}$O, $^{20}$Ne, $^{24}$Mg, $^{28}$Si,
$^{32}$S,  $^{40}$Ca) dominate the composition of the 
various layers.

The Fe core collapses
in timescales of milliseconds and bounces when nuclear densities are reached 
in its
inner regions. The resulting shock wave propagates outwards, losing
energy as it photo-disintegrates the onfalling outer Fe layers.
Numerical simulations show that the weakened shock  fails in general
to expel the stellar envelope  and a {\it 
prompt} explosion is not obtained. 
Neutrinos diffusing out of the
neutronized core on timescales of 0.1 sec and transferring part of their
energy and momenta to the outer Fe layers may lead to a successful
{\it delayed} explosion (e.g. Janka 1999 and references therein). 
In the meantime, the reverse shock produces
some accretion onto the proto-neutron star which, even after a
successful explosion is launched, may collapse to a black hole
(depending on its final mass).

The propagation of the shock wave in the stellar envelope heats the 
inner layers at temperatures appropriate for explosive Si 
($T_9 = T/10^9$ K $\simeq 4$), O ($T_9 \simeq 3.2$) and Ne/C ($T_9 \simeq 2$)
burning; due to the extended envelope structure, the outer He and H
layers never reach ignition temperatures. Explosive nucleosynthesis
in the O and Ne layers modifies somehow their pre-explosive abundance
pattern. In the Si layers, explosive burning accurs in two different regimes:
i) high density and low entropy and ii) high entropy and low density,
leading, respectively to a {\it normal} and {\it alpha-rich} ``freeze-out''
of nuclear reactions. In the former case the final abundances are in full
Nuclear Statistical Equilibrium (NSE). 
In the latter, some heavy Fe-peak nuclei are
also produced ($^{58}$Ni, $^{60}$Zn, $^{61}$Zn) and some $\alpha$-nuclei
are found in the final composition ($^{32}$S, $^{40}$Ca, $^{44}$Ti, $^{48}$Ca).
The most important of the Fe-peak nuclei produced in explosive Si-burning 
is $^{56}$Ni; its subsequent radioactive decay leads to the
production of $^{56}$Fe.  The story of the discovery of the ``radiogenic''
origin of $^{56}$Fe and its overall impact on Nuclear Astrophysics is
masterly described in the recent paper by D.D. Clayton (1999).

The resulting yields of the various isotopes 
are affected by the combined uncertainties  
of the input physics entering the pre-supernova evolution and the
explosion itself. The  uncertainties in 
{\it experimental reaction rates} used for
nuclei with A$<$30,  are evaluated in the recent compilation of the
NACRE project (Angulo et al. 1999). Their impact on hydrostatic H- and
He- burning nucleosynthesis is explored in Arnould et al. (1999). The
most important of these uncertainties concerns the
$^{12}$C($\alpha,\gamma$)$^{16}$O reaction; 
its combined effect with the mixing processes  determines not only the
final C/O ratio, but also the ratio of C-burning/O-burning products
as well as the size of the Fe core (Weaver \& Woosley 1993).
Despite the considerable amount of work devoted to the study of that
particular reaction, its rate is still uncertain by a factor of $\sim$2 at
He-burning temperatures.
For nuclei  with A$>$30 or for unstable ones (involved in explosive burning)
{\it theoretical reaction rates} are used.
A recent analysis (Hoffman et al. 1999) showed that, because chemical 
equilibria are attained in explosive burning,  the dependence of the yields
on cross section uncertainties is rather small for explosive O- and Si-burning
(less than $\sim$20\%).

Core {\it convection} in massive stars is usually treated either with  the
Schwarzchild criterion or with the Ledoux criterion, the latter leading
to somewhat smaller convective cores. 
Since the final yield depends on both core size and relevant nuclear reaction 
rates, it is impossible to constrain independently 
each one of those input physics by 
comparing theory to observations. Besides, all detailed nucleosynthesis 
calculations have been done up to now with 1-D codes neglecting the effects of
rotation. ``New generation'' models, including the effects of rotational
mixing show that the convective core is larger w.r.t. the non-rotating one;
the largest differences are found for H and He cores and the yields of the
corresponding burning phases  (Heger et al. 1999).
Moreover, rotational
mixing may lead to the production of {\it primary} N in massive stars, thus
helping  to solve
an  old ``puzzle'' concerning 
the galactic chemical evolution of nitrogen (see Sec. 4).
On the other hand, 2-D hydrodynamical simulations of O-shell burning by 
Bazan and Arnett (1998) revealed a complexe regime of
convective instabilities that 1-D models cannot describe adequately, making
the authors to ``view with scepticism the results of 1-D simulations at
that stage...''.
Clearly, the treatment of various mixing processes is still the single most
important problem in stellar astrophysics, and it affects considerably the
stellar yields.

{\it Mass loss} is another factor affecting (indirectly) 
the size of the convective core
and, ultimately, the stellar yields. It depends on both stellar mass and
metallicity, since radiation pressure on the envelope depends on the 
temperature of the radiation field and the abundance of metallic ions. For
metallicities Z$<$Z$_{\odot}$/20 mass loss has presumably 
a negligible effect on the
yields of stars of all masses. For Z=Z$_{\odot}$, stars with M$>$35 M$_{\odot}$
have the largest part of their envelope expelled before the formation of
the Fe-core. In that case, stars reach the Wolf-Rayet (WR) stage and
release through their stellar winds large amounts of H- and He- burning
products, in particular $^4$He, $^{14}$N and $^{12}$C
(Maeder 1992); these products may have some impact on the 
galactic evolution of C and N (see Sec. 4).
Up to now, very few  self-consistent calculations 
including mass loss have followed  all the burning stages,
up to the Fe-core formation and the subsequent explosion; in particular, 
Woosley et al. (1993, 1995) have shown the impact of mass
loss on the pre-supernova structure. However, it should
be emphasized that mass loss is still poorly understood, especially in the case
of WR stars, and adopted empirical prescriptions are uncertain by, at
least a factor of 2. This uncertainty is also reflected in the resulting
yields, in particular those from H- and (early) He-burning.

The {\it initial metallicity} of the star affects not only mass loss, 
but also the
outcome of nucleosynthesis. During H-burning the initial CNO transforms to 
$^{14}$N, and  part of the latter nucleus 
turns into $^{22}$Ne during He-burning
(through $\alpha$ captures and one $\beta$ decay). $^{12}$C, $^{14}$N and
$^{16}$O all have equal numbers of neutrons and protons
but not $^{22}$Ne (10 protons and 12 neutrons). This suprlus of neutrons 
(increasing with initial metallicity) affects the products of 
subsequent burning stages and, in particular, of explosive 
burning, favouring the production of odd nuclei ({\it ``odd-even''} effect).

The calculation of the Fe-core collapse {\it supernova explosion} is still one
of the major challenges in stellar astrophysics. 
Multi-dimensional hydrodynamical simulations in the 90ies revealed the
crucial role played by neutrino transport in the outcome of the explosion
(see  Janka 1999 and references therein). In the absence of a well-defined
explosion scheme, modelers of supernova nucleosynthesis
have to initiate the explosion somehow (by introducing either an ``internal
energy bomb'', or  a ``piston'', e.g. Aufderheide et al. 1991) 
and adjust the shock energy as to have a pre-determined 
final kinetic energy, usually the ``classical'' value of 10$^{51}$ ergs 
(after accounting for the binding energy of the ejected matter).
This procedure introduces one more degree of uncertainty in
the final yields. Moreover, the ejected amount of   Fe-peak nuclei
depends largely on the position of the {\it mass-cut}, the surface
separating the material falling back onto the neutronized core from the
ejected envelope. The position of this surface depends on the details
of the explosion (i.e. the delay between the bounce and the
neutrino-assisted explosion, during which the proto-neutron star
accretes material) and cannot be evaluated currently with precision
(see Thielemann et al. 1999 and references therein).

On the basis of energetic arguments, it seems plausible that
the explosion fails in the case of the most massive stars, which collapse
to form {\it black holes}; but,  even if
a successful explosion is launched,  the inner layers of the most
massive stars may also collapse shortly afterwards 
to a black hole (e.g. Freyer 1999 and 
references therein). This collapse, trapping the heavy elements inside the
compact object, may certainly affect the various abundance ratios
in the ejecta. However, neither the minimum initial  mass of those stars,
neither the final black hole masses can be reliably calculated  
at present. It should be stressed that, contrary to widespread views, 
nucleosynthesis arguments alone cannot determine the mass limit for stellar
black hole formation, because of the many uncertainties still affecting the
yields (Prantzos 1994).

In the light of the above, 
intermediate mass elements produced in massive stars may be 
divided in three main classes: \par
- In the 1st class belong N, C, O, Ne and Mg, which are mainly produced 
in hydrostatic burning phases and are found mainly in layers that are not
heavily processed by explosive nucleosynthesis; the yields of those elements
depend on the pre-supernova model (convection criterion, mixing processes,
mass loss and nuclear reaction rates). \par
- In the 2nd class belong Al, Si, S, Ar and Ca. 
They are produced by hydrostatic
burning, but their abundances are substantially affected by the passage of 
the shock wave. Their yields depend on both the pre-supernova model and the
shock wave energy. \par
- In the 3d class belong the Fe-peak nuclei, as well as some lighter elements
like Ti; their yields depend crucially upon the explosion mechanism and the
position of the ``mass-cut''.  
 
\begin{figure*}
{\psfig{file=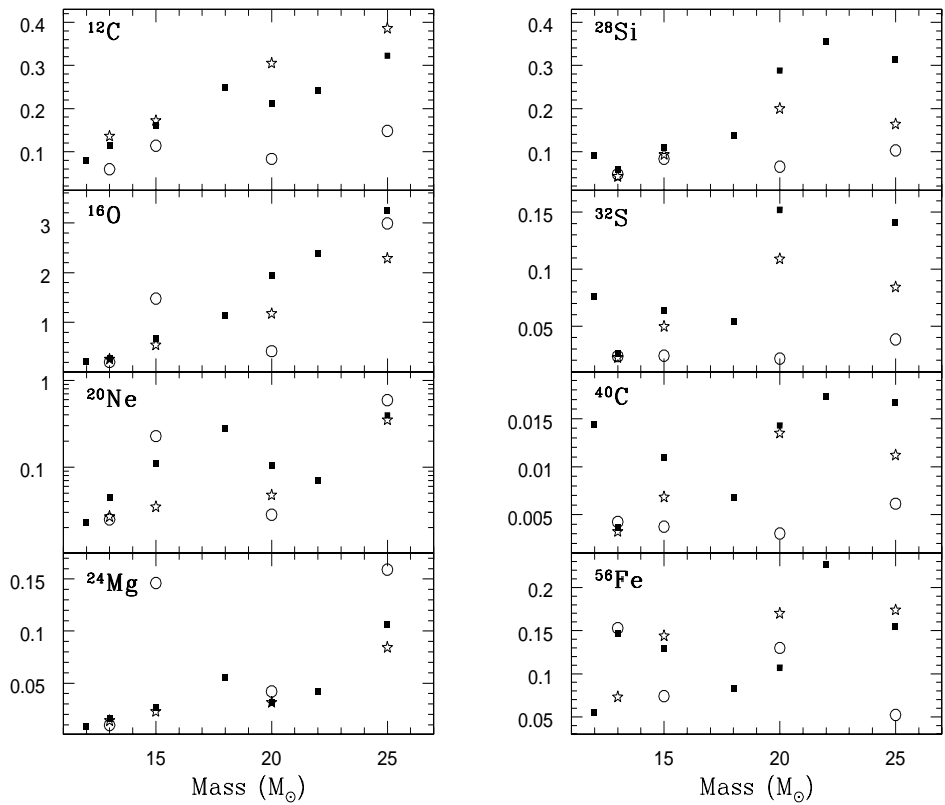,height=11.5truecm,width=\textwidth,angle=-90}}
\caption{\footnotesize 
Yields $Y(M)$ (ejected mass, in M$_{\odot}$) 
of several key elements as a function of initial stellar mass,
according to calculations for stars with solar initial metallicity.
{\it Filled squares}: Woosley and Weaver 1995; {\it Circles}: Thielemann
et al. 1996; {\it Asteriscs}: Limongi et al. 1999.
The spread between models at a given mass gives a rough idea of current
uncertainties. Notice that the yields do not show a monotonic behaviour with
mass.}
\end{figure*}

In the past five years or so 4 different groups have reported  results of 
(pre- and post-explosive) nucleosynthesis calculations in massive stars with
detailed networks. Thielemann et al. (1996) used bare He cores of initial
metallicity \Zs, while Arnett (1996) simulated the evolution of He cores 
(with polytropic-like trajectories) and studied different initial 
metallicities.
Full  stellar models (neglecting, however, rotation and  mass loss)
were studied by Woosley and Weaver 1995 (for masses 12, 13, 15, 18, 20, 22,
25, 30, 35 and 40 M$_{\odot}$ and metallicities Z=0, 10$^{-4}$, 10$^{-2}$,
10$^{-1}$ and 1 \Zs) and Limongi, Chieffi  and Straniero 1999
(for masses 13, 15, 20, 25 M$_{\odot}$ and metallicities  Z=0, 5 10$^{-2}$
and 1 \Zs        ). A critical analysis of the results of the former three
calculations has been done in the review of Arnett (1995).

A comparison of the yields of a few selected  elements in three of 
the aforementionned calculations is presented in Fig. 1. The yields of
C, O, Ne, Mg, Si, S, Ca and Fe are presented as a function of stellar
mass for stars with solar initial metallicities.  
The spread between models at a given mass gives a rough idea of current
uncertainties. Notice that the yields do not show a monotonic behaviour with
mass; this is true, in particular, for the Fe yields, which are the most
uncertain of all.

How can the validity of the theoretical stellar yields be checked?
Ideally, individual yields should be compared to abundances measured in 
supernova remnants of stars with known initial mass and metallicity!
However, such 
opportunities are extremely  rare. In the case of SN1987A, theoretical 
predictions for a 20 \Ms \ progenitor are in rather good agreement 
with observations of C, O, Si, Cl and Ar (Thielemann et al. 1996). 
SN1987A allowed also to ``calibrate'' the Fe yield 
($\sim$0.07 \Ms) from the optical light curve (powered at late times from
the decay of $^{56}$Co, the progeny of $^{56}$Ni), extrapolated
to the moment of the explosion (e.g. Arnett et al. 1989). 
This may be the best way to evaluate 
the Fe yields of other SNII at present,
until a convincing way of 
determining the ``mass-cut'' from first principles emerges.

\begin{figure*}
{\psfig{file=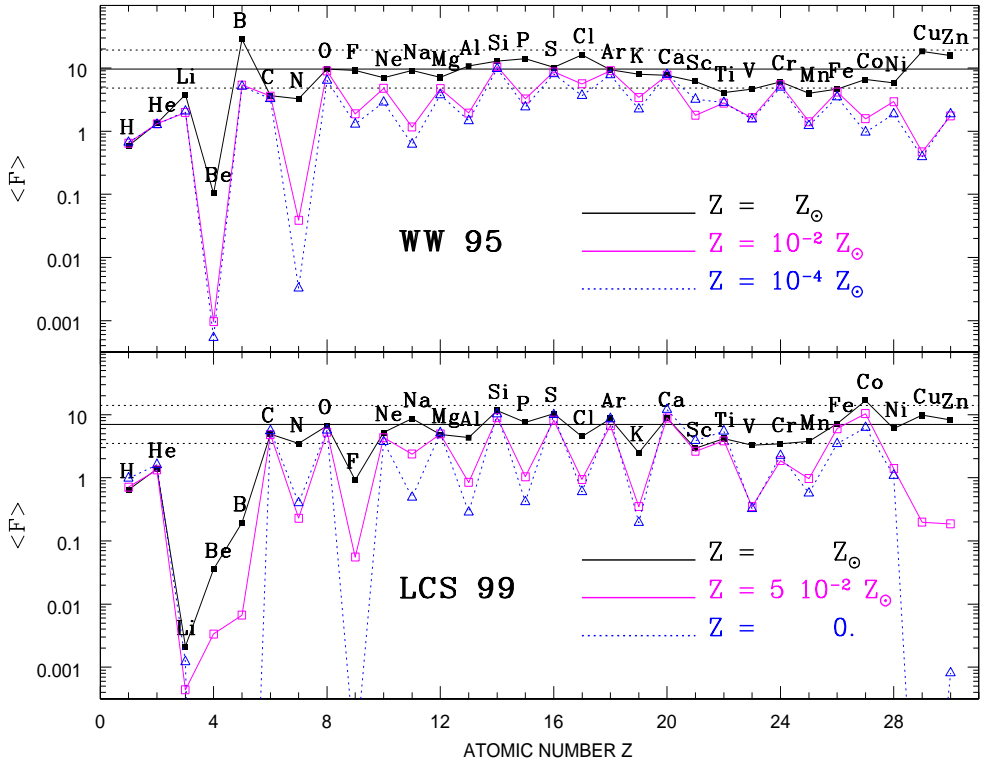,height=11truecm,width=\textwidth,angle=-90}}
\caption{\footnotesize Average overproduction factors 
(over a Scalo stellar IMF) of the yields of Woosley and Weaver 1995 (WW95, 
{\it upper panel}) and
Limongi, Chieffi and Straniero 1999 (LCS99, {\it lower panel}) 
for 3 different initial stellar metallicities.
In both cases, the {\it solid horizontal lines} are placed at $F_{Oxygen}$
and the two {\it dashed horizontal lines} at half and twice that value,
respectively. The ``odd-even effect'' is clearly seen in both data sets,
while N behaves as a pure ``secondary''.
The large values for Li, B and F in WW95 are due to $\nu$-process,
not included in LCS99. The elements
He, N, Li and Be in both cases (as well as B and F
in LCS99) obviously require another production site. This is also the case for
C in WW95.}
\end{figure*}

Finally, notice that
the overall yield used in chemical evolution studies depends on both
the individual stellar yields {\it and} the stellar IMF. 
Despite a vast amount of theoretical and observational work, 
the exact shape of the IMF is not yet well known
 (e.g. Gilmore et al. 1998 and references therein); it is clear, however,
that the IMF flattens in the low mass range and cannot be represented
by a power law of a single  slope. 
Our poor knowledge of the IMF
 introduces a further uncertainty of a factor $\sim$2 as to the
absolute yield value of each isotope (e.g. Wang and Silk 1993).

In Fig. 2 we present the yields  of WW95 and LCS99, folded with a Scalo 
(1986) IMF $\Phi(M)$.
They are compared to the corresponding mass of each isotope initially
present in the part of the star that is finally ejected.
The average oveproduction factor 
$F=\frac{\int Y(M)\Phi(M)dM }{\int X_{\odot}\Phi(M)(M-M_R)dM}$
(where $Y(M)$ is the yield and $M_R$ the mass of the remnant) 
is $F\sim$10 in the case of WW95 yields
with \Zs, i.e.
for each  gram of oxygen initially present in a stellar generation
10 grams  of oxygen are ejected (the remnant
mass should be properly subtracted).
It can be seen that most of the elements between O and Zn 
are nicely co-produced
(within a factor of 2), at least for stars with Z=\Zs.
Taking into account that the stars that contributed mostly to the
solar composition 4.5 Gyr ago had metallicities in the range 0.1 \Zs \ to
\Zs, Fig. 1 reveals that  the elements Sc, V and Ti are 
expected to be undeproduced
by the yields of both WW95 and LCS99 (see also Figs. 3 and 4); 
in fact, this is a feature shared by  the other
calculations of massive star nucleosynthesis.
It also transpires from Fig. 1 that He, C and N require 
another production site.

\section {Solar Neighbourhood: Absolute Yields and the Role of SNIa}

The composition of the proto-solar nebula (reflected in the well known
abundances of the solar photosphere and in meteorites, e.g. 
Anders and Grevesse 1989, Grevesse et al. 1996) is believed to be
representative of the local ISM 4.5 Gyr ago. It is difficult to know
to what extent this assumption holds true. Young stars and gas in the
nearby Orion nebula show CNO abundances smaller by $\sim$30\% than the
corresponding solar values (Cunha \& Lambert 1994). 
On the other hand, solar type stars of
similar ages show rather similar compositions (the scatter in the
age-metallicity relation of Edvardsson et al. 1993 is probably due to
contamination of their sample by stars from different galactic regions,
e.g. Garnett and Kobulnicky 1999). Thus, it is probably safe
to say at present that the above assumption is true within 30\%.

The proto-solar composition is the result of $\sim$10 Gyr of galactic
evolution, with succeeding generations of different type stars adding
their specific contribution in the galactic ``blender''. Since massive
stars are the major nucleosynthetic site of intermediate mass nuclei
(from O to Ge), it is clear that the theoretical yields discussed in
Sec.  2 should reproduce (within a ``reasonable'' factor) the solar
composition.
A full galactic chemical evolution model
of  the solar neighbourhood should be used for that exercise,
 since the physics of the model (infall, outflow, IMF, etc.)
affect in different ways the different species: {\it primaries} (with yields 
independent of initial stellar metallicity, like O), 
{\it secondaries} (with yields
proportional to metallicity, like N in current models of massive stars) 
or others (i.e. ``odd-even'' isotopes, with a mild yield dependence on 
metallicity, like Al).

At this point it should be noticed that there is 
a strong observational argument, suggesting that
massive stars {\it are not } the sole producers 
of Fe peak nuclei in the solar neighborhood. This stems from the 
observed decline  in the O/Fe ratio (see Fig. 4),
from its value of $\sim$3 times solar in halo stars 
([O/Fe] $\simeq 0.5$ for [Fe/H] $< -1$) to
solar in disk stars ([O/Fe] $\simeq -0.5$ [Fe/H] for $-1 <$ [Fe/H] $< 0$). 
This decline is usually interpreted
in terms of some late source of Fe (and Fe group  elements, since their
ratio to Fe remains about constant during the disk phase).
Taking into account that massive stars (assumed to be the only source
of O and Fe in the halo phase) produce  a Fe/O ratio $\sim$1/3 solar, 
the remaining $\sim$2/3 should be produced by that late source,
presumably SNIa. 
[{\it Notice:} this ``traditional'' view of O/Fe behaviour is challenged by
recent observations
showing a continuous decline of O/Fe, from the lowest halo metallicities
down to solar (Israelian et al. 1998, Boesgaard et al. 1999);
 these findings are not confirmed by other studies - Fullbright and Kraft
1999 - but the situation is still not clear. 
If the new findings are confirmed, the oxygen yields should be revised
and some dependence on metallicity introduced, probably due to mass
loss; the yields of other $\alpha$-elements, produced in inner stellar layers,
should be left unaffected by that revision. Here we stick to the ``old''
paradigm, i.e. O/Fe$\sim$constant in halo stars].

The best current model for SNIa nucleosynthesis
is believed to be the carbon deflagration
model W7 of Thielemann et al. (1986).
The deflagration, starting in the center of an
accreting Chandrasekhar mass CO white dwarf, burns about half of the stellar
material in NSE and produces 
$\sim$ 0.7 \Ms \ of $^{56}$Fe (originally in the form of $^{56}$Ni). 
It also produces all other Fe-peak isotopes and, in particular, $^{58}$Ni
(see below).

The problem with SNIa is
that, although the current rate of SNIa/SNII is 
constrained by observations in external spiral galaxies
(Tammann et al. 1994), the past
history of that rate (depending on the nature of progenitor 
systems) is virtually unknown. Thus, at present,
it is rather a mystery why the timescale for the onset of SNIa
activity (presumably producing the observed decline of O/Fe in the disk) 
coincides with 
the timescale for halo formation. Modelers  of galactic chemical evolution
circumvent this ``paradox'' by simply adjusting the timescale of SNIa 
progenitors (or the corresponding mass range), such as to make them 
``effective'' {\it after}  the halo phase. An original suggestion was 
recently made by Kobayashi  et al. (1998): in a system
composed of a white dwarf + an evolved companion,
the accreting white dwarf blows a wind which, if the metallicity is 
sufficiently high ([Fe/H]$>$-1), maintains a quasi-steady accretion rate and
allows the white dwarf to reach the Chandrasekhar mass and explode as SNIa
(curiously, the mechanism does not operate at lower metallicities).
The interest of this scenario lies in the fact that 
 SNIa enter the cosmic scene at just the right moment.

\begin{figure}
\psfig{file=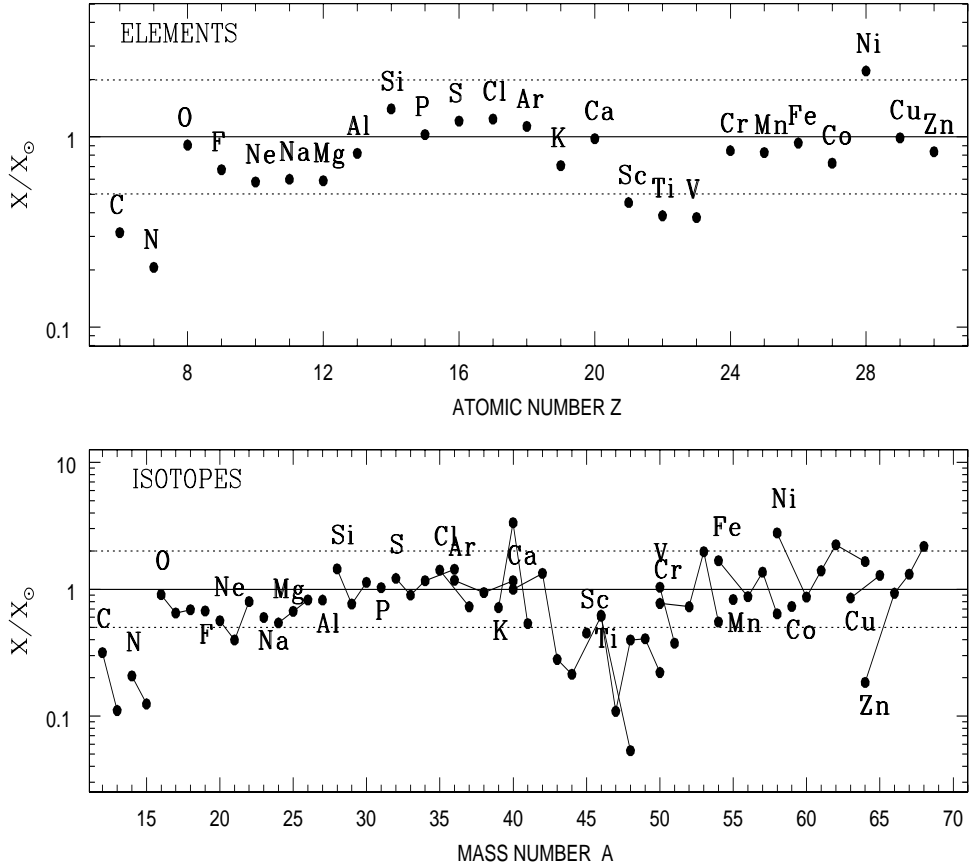,height=13.truecm,width=\textwidth,angle=-90}
\caption{\small 
Composition of the solar neighborhood 4.5 Gyr ago, obtained with a chemical
evolution model which reproduces all available local observational constraints
(current amounts of gas and stars, age-metallicity relation and G-dwarf
metallicity distribution) and utilises the WW95 metallicity-dependent yields
for massive stars and the W7 model of Thielemann et al. (1986) for SNIa.
C and N isotopes require another source (intermediate mass stars? Wolf-Rayet
winds?), not included here. The undeproduction of Sc, Ti, V is a common 
feature of all currently available sets of massive star yields. 
The overproduction of Ni (in the form of the main isotope $^{58}$Ni) results
from the W7 model of Thielemann et al. (1986) for SNIa.
}
\end{figure}

A few calculations have been performed up to now with the full set of 
metallicity dependent yields of WW95
(Timmes et al. 1995, Samland 1998,
while Thomas et al. 1998 considered only a few selected elements).
In Fig. 3 we show the results of a recent calculation (Goswami and Prantzos 
2000), using the WW95 metallicity dependent yields of massive stars and the 
W7 model for SNIa (yields of intermediate mass stars are not included);
the model reproduces all the currently available constraints in the solar
neighborhood. It can be seen that \par
- i) most elements and isotopes between
O and Zn are nicely co-produced (within a factor of two), \par
- ii) C and N require
another source (intermediate mass stars? WR stars?), \par
- iii) Sc, V and Ti are
slightly undeproduced, due apparently to the inadequacy of all
currently available sets of stellar yields (see Sec. 2), \par
- iv) there is a small
oveproduction of Ni, due to the isotope $^{58}$Ni, which is abundantly produced
in the W7 model of SNIa. The amount of $^{58}$Ni depends mostly on the central
density of the exploding white dwarf and the overproduction problem
may be fixed by varying this parameter; indeed, alternatives to the W7 model
have recently been calculated (Iwamoto et al. 1999).

Notice that for the calculation reported in Fig. 3, the Fe yields
of WW95 have been reduced by a factor of two, in order to reproduce
the observed O/Fe ratio in halo stars ($\sim$3 times solar, see Sec. 4); 
otherwise, the WW95 massive
stars alone can make almost the full solar abundance of Fe-peak nuclei (as
shown in Timmes et al. 1995), leaving no room for SNIa. Taking into account
the uncertainties in the yields, especially those of Fe-peak nuclei (see
Sec. 2) our reduction imposed on the WW95 Fe yields is not unrealistic.

The nice agreement between theory and observations in Fig. 3 comes as
a pleasant surprise, in view of the many uncertainties discussed in the
previous section. It certainly does not guarantee that each individual
yield is correctly evaluated. It rather suggests that the various factors
of uncertainty  cancel out (indeed, it is improbable that they all
``push'' towards the same direction!)  
so that an overall satisfactory agreement with observations results. 
Thus, at least to first order, one may say that 
currently available yields of massive stars + SNIa can account for the solar 
composition between O and Zn (baring Sc, Ti and V).

\section {Galactic Halo: Yield Ratios and Earliest Nucleosynthesis}

Observations of metal abundance ratios in long-lived (M$\simeq$1 \Ms)
main sequence stars of  $Z <$ \Zs \ offer invaluable information 
about the past history of nucleosynthesis in our Galaxy
(e.g. Matteucci 1996, Pagel 1997).

In particular, the X/Fe ratio as a function of metallicity gives
information about: 

i) the mass of the progenitor star of species X, through
the delay introduced by the corresponding finite stellar lifetime
(i.e. the products of intermediate mass stars
appear later than those of massive stars);

ii) the  properties of binary systems, through the delay of e.g. 
Fe production in SNIa;

iii) any metallicity dependence in the yield of X: primary vs. secondary species, ``odd-even'' effect,
or simply metal-dependent stellar mass loss (as for C and N in 
in the $M > 40$ \Ms \ stars of Maeder 1992, see also Sec. 2).

\begin{figure}
\psfig{file=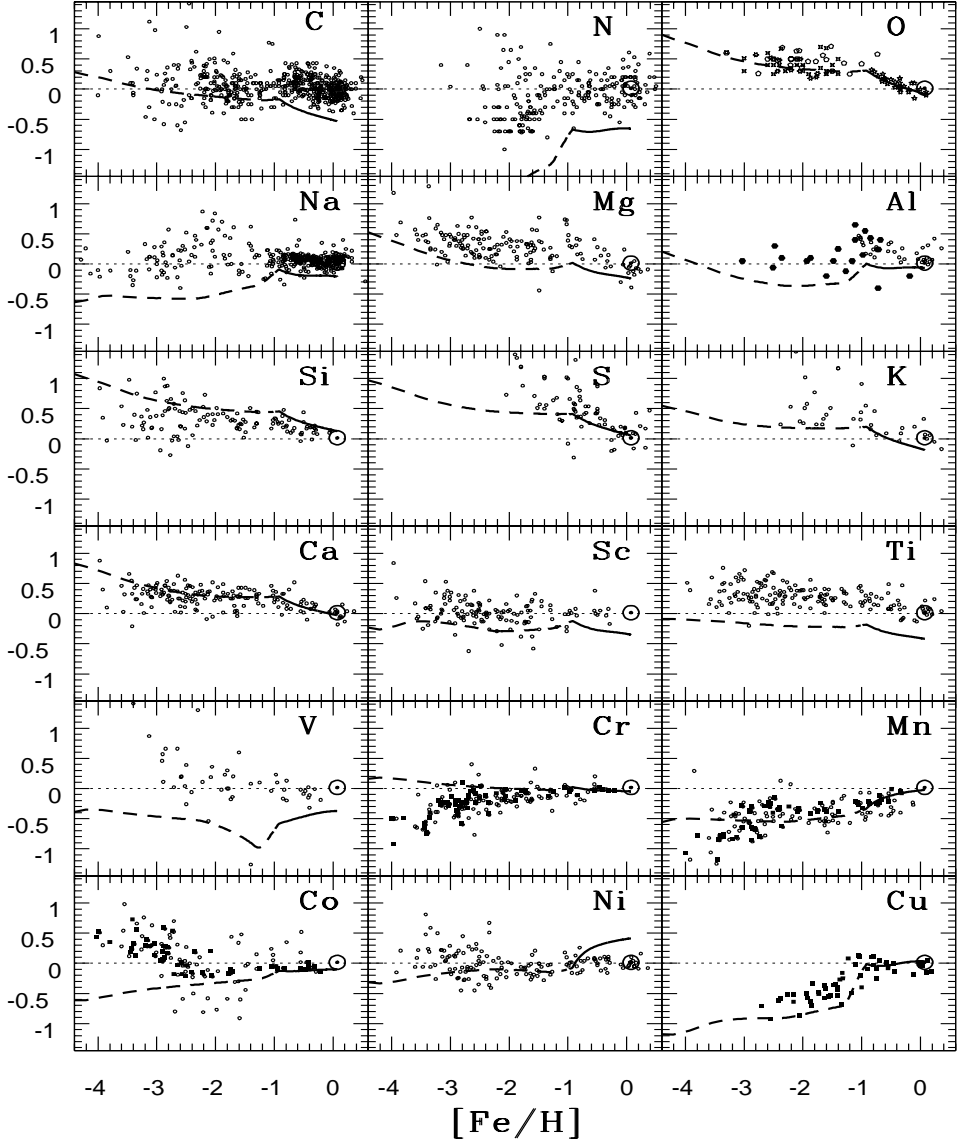,height=16.5cm,width=\textwidth}
\caption{\small Abundance ratios [X/Fe] of stars in the halo and the
local disk, as a function of [Fe/H].
They are compared to the results of a 
detailed chemical evolution model (Goswami and Prantzos
2000), utilising the
metallicity-dependent yields of WW95 for massive stars and the W7 model
for SNIa; yields from intermediate mass stars are {\it not} included.
The model treats properly the halo ({\it dashed curve} assuming outflow) 
and the
disk ({\it solid curve} assuming slow infall), 
in order to reproduce the corresponding 
metallicity distributions of low mass stars. Notice that the WW95  yields
of Fe have been divided by 2, in order to obtain the observed $\alpha$/Fe ratio
in halo stars. Model trends below [Fe/H]=-3 are due to the finite lifetime of
stars ([Fe/H]=-4 is attained at 10 Myr, i.e. stars of $>$20 M$_{\odot}$
have time to explode, while  
[Fe/H]=-3 is attained at 20 Myr, i.e. stars of $>$10 M$_{\odot}$ explode);
in view of the  yield uncertainties in individual stars, 
{\it they should not be considered
as significant}.}
\end{figure}

Observations of various abundance ratios in low-metallicity
stars (e.g. Ryan et al. 1996,  McWilliam 1997, for recent major surveys)  
allowed to establish several trends (Fig. 4). 
Several works in the past few years attempted to understand some (or most of)
these trends (e.g. Chappini et al. 1999).
In particular, those of Timmes et al. (1995) and Samland (1998)
are the most comprehensive, 
surveying the full range of elements from C to Zn
and using the metallicity dependent yields of WW95. Their models present
some differences: Timmes et al. (1995) use a simple model with infall (in fact,
appropriate only for the disk, but certainly not for the halo!), and  consider
the full range of stellar masses and corresponding lifetimes; Samland (1998)
uses a full dynamical model (treating, presumably, correctly, the halo and
the disk) but makes a few approximations concerning the stellar lifetimes and
the metallicity dependence of the yields.
The conclusions of both works are basically similar to those of 
the recent work of Goswami and Prantzos (2000). The latter utilises
simplified appropriate models for the halo (with outflow) and the disk 
(with slow infall) as to reproduce all currently available constraints and,
in particular, the corresponding metallicity distributions of G-dwarf stars.
The halo and the disk are treated as independent systems, not connected
by any temporal sequence. The disk starts with essentially zero initial
metallicity and very small amount of gas; the number of stars formed in the
disk  at 
[Fe/H]$<$-1 is negligible. In Fig. 4 we plot our results for both
the halo ({\it dashed curve}) and the disk ({\it solid curve}, 
only stars with [Fe/H]$>$-1, which constitute the vast majority). Notice that,
in order to evaluate  the impact of the yields of massive stars,
we do not include any yields from intermediate or low masss stars or novae.
Also, notice that the WW95 yields of Fe are divided by 2, in order to 
reproduce the observed O/Fe$\sim$3 time solar in halo stars.
It can be seen that: 

- The observed C/Fe$\sim$const.  in the disk is not reproduced; some late
C source is required, 
either from long lived, low mass stars, or, most probably, from
metallicity-dependent Wolf-Rayet winds (Prantzos et al. 1994, 
Gustaffson et al. 1999) 

- An early  source of {\it primary N} in the halo is required, in order to
reproduce the observed N/Fe$\sim$constant ratio. As discussed in Sec. 2,
curent massive star models produce only secondary N, through the CNO cycle.
Mixing of protons (induced by rotation) in the He-burning core, could
lead to the production of primary N in those stars (by p-captures on 
$^{12}$C, itself produced by He-burning) and thus help to solve the puzzle
(G. Meynet, private communication).

- Oxygen  and the other $\alpha$-elements (Mg, Si, S, Ca, Ti) show a similar
behaviour (if we neglect the recent findings of Israelian et al. 1998 and
Boesgaard et al. 1999 for O, see Sec. 3). The observed trend is readily
understood in terms of SNII contribution in the halo and SNII+SNIa in the 
disk (e.g. Pagel and Trautvaisienne 1995). However, the WW95 yields of Mg
are low w.r.t. the observations of halo stars, as already noticed by several 
authors (e.g. Timmes et al. 1995, Thomas et al. 1998).

- The observed Na/Fe and Al/Fe ratios do not exhibit the expected ``odd-even''
behaviour; instead, they seem to behave like pure primaries.

- Sc, V and Ti are systematically undeproduced w.r.t. Fe at all metallicities
as already noticed by Timmes et al. (1995).

- Among Fe peak elements, only  Mn  shows the theoretically expected
``odd-even'' behaviour; Copper, produced mostly in hydrostatic nucleosynthesis,
shows a similar behaviour, well reproduced by theory.

- Cr, Co and Ni behave in a rather unexpected way at low metallicities:
Cr and Ni are expected to follow Fe, but this does not seem to be the case;
Co/Fe should be lower than solar at low metallicities, but the opposite is
observed. Nakamura et al. (1999) tried to interpret these data by varying
the mass-cut of the Thielemann et al. (1996) models (obtained for solar
metallicity stars); they had some success concerning Cr and Ni, but the case
of Co remains difficult to understand. 

Notice that the ``trends'' of our model  below [Fe/H]=-3 are due to the 
finite lifetime of stars: [Fe/H]=-4 is attained at 10 Myr, i.e. stars of 
$>$20 M$_{\odot}$ have time to evolve and explode, while  
[Fe/H]=-3 is attained at 20 Myr, i.e. stars of $>$10 M$_{\odot}$ explode;
in view of the  yield uncertainties in individual stars, 
{\it these  model trends should not be considered
as significant}. Moreover, 
these early times of galactic history are characterised
by composition inhomogeneities: the gas is contaminated by the ejecta of only
a few supernovae, since the mixing timescales are comparable to galactic
evolution timescales. Our model, with its instantaneous mixing 
approximation, cannot apply in such conditions (this is also the case for
all ``classical'' models of galactic chemical evolution;
models treating those inhomogeneities have also been developed, 
but they introduce at least one more free parameter....)

\section {Conclusions}

The nucleosynthetic yields of massive stars are still subject to important
theoretical uncertainties (due to our poor understanding of: mass loss, 
convection and mixing in general, explosion mechanism and
several key  nuclear reaction rates). To a first approximation, our
current understanding of massive star nucleosynthesis seems sufficient to
explain the solar system composition between O and Zn (with the exception
of Sc, Ti and V), and the abundance patterns of some (but not all!)
$\alpha$-elements in halo stars. The yields of Fe-peak nuclei (being
extremely sensitive to the explosion mechanism) remain highly uncertain at
present. Observations of individual events (i.e. abundances in supernova
remnants) and of very low metallicity stars could contribute towards some
progress in that direction.

{}
\end{document}